\documentclass{article}
\usepackage[utf8]{inputenc}
\usepackage{authblk}

\usepackage[margin=1.2in]{geometry}
\title{A comparative study of forecasting Corporate Credit Ratings using Neural Networks, Support Vector Machines, and Decision Trees}
\author[1]{Parisa Golbayani} 
\author[2]{Ionu\c{t} Florescu}
\author[3]{Rupak Chatterjee}
\affil[1]{Financial Engineering, School of Business, Stevens Institute of Technology}
\affil[2]{Hanlon Financial Systems Center,  Financial Engineering, School of Business, Stevens Institute of Technology}
\affil[3]{Department of Physics, Center for Quantum Science and Engineering, Schaefer School of Engineering and Science, Stevens Institute of Technology}
\usepackage{setspace}
\usepackage{natbib}
\usepackage{graphicx}
\usepackage{amsmath}
\usepackage{amsthm}
\usepackage{float}
\usepackage{adjustbox}
\usepackage{subcaption}
\usepackage[colorlinks=true,linkcolor=blue,citecolor = blue]{hyperref}%

\usepackage[usenames,dvipsnames]{xcolor}

\newtheorem{theorem}{Theorem}

\theoremstyle{definition}
\newtheorem{definition}{Definition}

\begin{document}

\maketitle

\section*{Abstract}
Credit ratings are one of the primary keys that reflect the level of riskiness and reliability of corporations to meet their financial obligations. Rating agencies tend to take extended periods of time to provide new ratings and update older ones. Therefore, credit scoring assessments using artificial intelligence has gained a lot of interest in recent years. Successful machine learning methods can provide rapid analysis of credit scores while updating older ones on a daily time scale. Related studies have shown that neural networks and support vector machines outperform other techniques by providing better prediction accuracy. The purpose of this paper is two fold. First, we provide a survey and a comparative analysis of results from literature applying machine learning techniques to predict credit rating. Second, we apply ourselves four machine learning techniques deemed useful from previous studies (Bagged Decision Trees, Random Forest, Support Vector Machine and Multilayer Perceptron) to the same datasets. We evaluate the results using a 10-fold cross validation technique. The results of the experiment for the datasets chosen show superior performance for decision tree based models. In addition to the conventional accuracy measure of classifiers, we introduce a measure of accuracy based on notches called "Notch Distance" to analyze the performance of the above classifiers in the specific context of credit rating. This measure tells us how far the predictions are from the true ratings. We further compare the performance of three major rating agencies, Standard $\&$ Poors, Moody's and Fitch where we show that the difference in their ratings is comparable with the decision tree prediction versus the actual rating on the test dataset.

\section{Introduction}
Machine learning models have been widely applied on a range of applications in financial fields due to their capability of detecting embedded trends in financial data. For example, asset prices follow a random, non linear dynamic pattern due to many factors such as political events, economic conditions and traders behavior \citep{huang2005forecasting}. Despite this non linear nature of price movements, machine learning techniques provide an accurate estimation of asset prices \citep{kim2003Financial, huang2005forecasting, kanchymalay2017multivariate, ganguli2017machine, culkin2017machine,feng2018deep,gu2018empirical, sambasivan2017statistical}. 
In addition to asset pricing, machine learning techniques gained a lot of interest in credit rating prediction problem. Credit ratings have been used as an informative indicators of the level of riskiness of companies and bonds \citep{huang2004credit, lee2007application}. They reflect the likelihood that a counterparty will default on his/her financial obligation \citep{martens2010credit}. Rating agencies such as Standard and Poor's and Moodys analyze various aspects of companies financial status to come up with these credit ratings \citep{huang2004credit}. Such an assessment is an expensive complicated process often taking months since it involves many experts to analyze different variables that reflect the underlying corporations' reliability \citep{hajek2013feature}. One solution to reduce this financial and time cost is to come up with a predictive quantitative model based on the historical financial information of a company.

In this study we provide a comparative analysis of the literature of Corporate credit rating prediction problem. We focus on modern statistical and Artificial Intelligence techniques. We review and summarize the performance of techniques that were demonstrated to be the most accurate in the literature. We implement these most accurate techniques and compare their performance when analyzing three different sectors of the US economy. To our knowledge, this is one of the first papers that implements corporate rating models on the same data and thus compares algorithms' performance on an equal footing. 

Corporate credit rating is a different problem than the typical image classification techniques. Thus, we introduce a new evaluation measure: the \textit{Notch Distance} in section \ref{sec:NotchDistance} on page \pageref{sec:NotchDistance}. This measure allows us to better quantify and distinguish between performance of various machine learning methods. 

This paper is organized as follows. In Section 2, we provide a review of prior literature on the credit rating problem. Section 3 contains an explanation of each machine learning method studied in order to predict corporate credit ratings. Section 4 explains the nature of the input and output data sets in this study as well as the evaluation methodology used. Experimental results and analysis are provided in Section 5. Final conclusions are drawn in section 6, together with future applications of the methodology. 

\section{Prior Methods on the Credit Rating Problem using Machine Learning Techniques} 

\paragraph{Bond Rating} The majority of literature on the credit rating problem is focused on predicting bond's probability of default and thus bond rating \citep{saha2017credit}. Garavaglia \citep{garavaglia1991application} simulated S\&P corporate bond rating using the unidirectional version of the counter-propagation neural network. This neural network model introduced by Robert Hetch-Nielsen. Chaveesuk et al. \citep{chaveesuk1999alternative} studied three supervised neural networks concepts, backpropagation, radial basis function and learning vector quantization. However, they concluded, neither a neural network nor a regression model performs well on bond rating assignments. Shin and Han  \citep{shin2001case} combined inductive learning approach based on decision trees with case indexing process in order to build a knowledge-based system that can classify Korean corporate bond ratings. Kim \& Han \citep{kim2001cluster} clustered financial data among bond rating groups. They used competitive artificial neural network to generate the centroid values of the clusters. They perform a case-indexing study based on these cluster information and combine it with the classification technique in order to present a hybrid model for forecasting corporate bond ratings. They apply their method on 167 financial variables including five categorical variable and 162 financial ratios. After performing some statistical tests such as ANOVA, factor analysis and step-wise method, they select 13 financial ratios as the final input data. Abdou \citep{abdou2009evaluation} studied feed forward neural network  to evaluate consumer loans within the Egyptian private banks. Lessman et al. \cite{lessmann2015benchmarking} provided a deep analysis of machine learning techniques used in forecasting retail credit scoring. They studied $41$ classification algorithms used to predict probability of default for retail loans. The authors used multiple datasets and criteria for evaluation and concluded that their hybrid method is superior to other methods.  Gangolf et al. \cite{gangolf2016automated} provided a comparison of statistical and artificial intelligence techniques used for forecasting probability of default for bonds. In case of predicting probability of default, Angelini et al \citep{angelini2008neural} studied two different neural networks to evaluate credit risk of Italian small businesses. They showed that neural network architectures can be useful for estimating the probability of default of borrowers. Tsai et al. \citep{tsai2008using} studied the application of multilayer perceptron network on bankruptcy prediction and credit risk evaluation. Their analysis proved that single neural network provides accurate results compared to multi-layers classifiers. Addo et al. \citep{addo2018credit} studied seven binary classifiers (elastic net, random forest, gradient boosting model and four different neural network architectures) in predicting loan default probability. They implemented these models on 117019 data points in the year 2016 and 2017 provided from European banks. They used $60\%$ of data as training set, $20\%$ as validation and $20\%$ as the test set. They compared the performance of these models on a selected subset of important features based on variable importance of each model. They concluded three important points. First, that decision tree based models outperform multilayer artificial neural networks. Second, that standard performance criteria such as AIC or R2 are not sufficient enough to compare different models. Third, that the selected subset of important features based on variable importance of models, does not necessarily provide stable results.

\paragraph{Corporate credit rating} In this work we are concerned with \textit{corporate credit rating}. Although, the two problems are related, the bond ratings are not necessarily the same as the issuing company's corporate rating due to the priority of claim of various bonds. There are many more bonds than there are corporations and thus the data availability is much more limited. When rating publicly traded corporations the input is based generally on the 10-Q  and 10-K statements. However, bonds are traded and thus bond values can be marked to market far more frequently than quarterly. Furthermore, while the basic bond structure is similar for all issuing companies regardless of sector, the quarterly statement issued by a bank is fundamentally different from that issued by a company from say the Consumer Staples sector. Thus papers dedicated to corporate credit rating are generally constructing models for companies from a specific sector. The lack of large data sets is thus typical of this domain and this makes the credit rating problem challenging.

Methodologically, statistical models and artificial intelligence approaches are used in the field of credit rating prediction. In this section, we provide a review of prior and relevant literature of models used. We start from standard models and later present more recent and complex models. We summarize these studies in Tables \ref{tab:1}, \ref{tab:2} and \ref{tab:3}. We are basing our subsequent analysis on the accuracy numbers in these tables to determine which models will be implemented on the financial statements data.

\paragraph{Statistical Methods}
Standard statistical techniques such as logistic regression analysis (LRA) and discriminant analysis (DA) have been used to construct credit scoring models \citep{horrigan1966determination, west1970alternative, kaplan1979statistical, kim1993expert,kamstra2001combining,oelerich2006evaluation,abdou2009evaluation,akkocc2012empirical,abdou2016predicting}. Modified version of these techniques have been developed to improve credit scoring models. \citep{hwang2009multiple,hwang2010predicting} modify the usual ordered probit model by replacing the linear regression function  with a semi-parametric function to increase the choices of regression function. They apply this prediction method on four market driven variables, nineteen accounting variables and industry indicator variables. Their conclusion shows that an ordered semiparametric probit model (OSPM) provides a better prediction than ordered linear probit model (OLPM). However, different assumptions behind most of these statistical approach, such as linearity, normality and  distribution-dependency limit their predictive capability comparing to Artificial Intelligence techniques \citep{huang2004credit,pai2015credit}. 

Nonetheless, recent papers are still using traditional statistics methods and new developments. \citet*{gogas2014forecasting} used ordered probit regression model to forecast long term bank credit rating. The benchmark was Fitch credit rating in 2012.  
\cite{irmatova2016relarm} introduced a new rating model based on principal component analysis and k-means clustering algorithm to predict $30$ country's rating. \cite{karminsky2016extended} implemented an ordered probit regression model to predict bank ratings using the Bankscope database. \cite{hajek2016predicting} developed a methodology based on latent semantic analysis to extract topical content from company-related documents. The authors use naive bayesian network to forecast corporate credit ratings. Recently, \citep{petropoulos2016novel} developed a methodology based on Student's-t Hidden Markov Models (SHMMs) to investigate the temporal dimension and heavy tailed distribution of firm-related data.

We next present three prevalent AI techniques used for corporate credit rating: Artificial Neural Networks (ANN), Support Vector Machines (SVM) and Classification and Regression Trees (CART). 

\paragraph{Artificial Neural Networks (ANN)}
Artificial Neural Networks draws attention from many researchers with its robustness, flexibility and higher accuracy \citep{dutta1988bond,surkan1990neural, kim1993expert, maher1997predicting,kwon1997ordinal,Brennan2004CorporateBR,abdou2009evaluation,abdou2016predicting}. West \citep{west2000neural} implemented five neural network architectures: multi-layer perceptron, mixture-of-experts, radial basis function, learning vector quantization and fuzzy adaptive resonance. He compared the results of these neural networks with linear discriminant analysis, logistic regression, k nearest neighbor, kernel density estimation, and decision trees. They studied German and Australian credit scoring data and conclude that multi-layer perceptron is not the most accurate neural network architecture and logistic regression outperforms all neural network architectures. Kim \citep{kim2005predicting} applied adaptive learning networks (ALN) which is a non-parametric model, on both financial and non financial variables to predict S\&P credit ratings. Yu et al. \citep{yu2008credit} proposed a six stage neural network ensemble learning model to assess a credit risk measurement on Japanese consumer credit card applications approval and UK corporations. Kashman \citep{khashman2010neural} investigated three neural networks based on back propagation learning algorithm on German Credit Approval dataset. The architecture of these neural networks are different according to various parameters used in the model such as hidden units, learning coefficients, momentum rate and random initial weight range. Pacelli and Azzollini \citep{pacelli2011artificial} provided an overview of different types of neural networks used in credit rating literature. 

\paragraph{Support Vector Machines}
Among all artificial intelligence techniques, Support Vector Machine (SVM) has indicated a powerful classification ability \citep{cortes1995support, kim2010support, vapnik2013nature,xiao2016ensemble}. However the standard SVM, addresses the two-class classification problems. For multi-class classification problems, SVM can be represented as several two class tasks such as one-against-one (OAO-SVM) \citep{scholkopf1999advances}, one-against-all (OAA-SVM) \citep{bottou1994comparison} and direct acyclic graph (DAG-SVM) \citep{platt2000large,wang2009novel}. Cao et al. \citep{cao2006bond} studied all these three approaches on US companies from the manufacturing sector. Hajek \& Olej \citep{hajek2011credit} studied Support Vector Machines with supervised learning as well as kernel-based approach with semi-supervised learning in order to predict corporate and municipal credit ratings. Kim \& Ahn \citep{kim2012corporate} studied a new type classifier based on Support Vector Machine, called (OMSVM) which applies ordinal partwise partitioning (OPP) to extend the binary SVM. 

\paragraph{Decision Trees}
According to the literature, some researchers focus on the higher level of interpretability and feature extraction flexibility of decision tree-based approach in multi-class classification problems \citep{lee2006mining,kao2012bayesian,abdou2016predicting}. The fundamental of these techniques are introduced first by \citep{breiman1984classification}.
Paleologo et al. \citep{paleologo2010subagging} proposed a missing data imputation method and ensemble classification technique, subagging, particularly for unbalanced credit rating data sets. The proposed subagging technique is based on different models such as decision trees and support vector machines. Khandani et al. \citep{khandani2010consumer} studied general classification and regression tree technique (CART) on a combination of traditional credit factors and consumer banking transactions to predict consumer credit risk.
Pedro Veronezi  \citep{pedro} applied Random Forest(RF) and Multi Layer Perceptron (MLP) to predict corporate credit ratings using their financial data. First, he used RF to perform feature selection and then the result was fed to two different RF models. One model aims to predict the changes and another model predicts the direction. The result of these two models have been added to the input dataset. After creating a new dataset, two models were trained: a MLP model and a RF model. He concluded that RF outperforms MLP by providing a more accurate and stable results in a shorter period of time.  

\paragraph{Hybrid Models}
In addition to all these frequently used techniques, some researchers study other approaches to provide a credit scoring model. Peng et al. \citep{peng2011empirical} introduced three multiple criteria decision making (MCDM) methods to evaluate classification algorithms for financial risk prediction. Chen \citep{chen2012classifying} investigated rough set theory (RST) approach to classify Asian banks' ratings. Some researchers take one step further and integrate multiple techniques to achieve a higher accuracy. Yeh et al. \citep{yeh2012hybrid} combined random forest feature selection with different approach such as rough set theory (RST) and SVM. They tested these techniques on 17 financial variables in addition to the market based information which has been obtained by Moody’s KMV tool for each corporation. Wu \& Hsu \citep{wu2012credit} introduced an enhanced decision support model which is a combination of relevance vector machine and decision tree. Hajek \citep{hajek2012credit} used fuzzy rule-based system adapted by a feed-forward neural network to classify credit ratings of US municipalities and companies. Pai et al. \citep{pai2015credit} used Decision Tree Support Vector Machine (DTSVM) integrated to multiple feature selection strategies and rough set theory (RST) to predict credit ratings. Description for other hybrid techniques are available in \citep{chandra2009failure,akkocc2012empirical,chen2013hybrid}.    

\paragraph{Evaluation and Comparison}
The Comparison of the discussed techniques with different parameters and architectures on several datasets, have been widely studied in credit scoring field. Huang et al. \citep{huang2004credit} studied support vector machines (SVM) and back-propagation neural network (BNN) to obtain more accurate credit rating predictions for Taiwan financial institutes and United States commercial banks. Lee \citep{lee2007application} used 5-fold cross-validation to find the optimal parameters of SVM, multiple discriminant analysis (MDA), case-based reasoning (CBR), and three-layer fully connected back-propagation neural networks (BPNs) on Korean Datasets. Ye et al. \citep{ye2008multiclass} compared the performance of four different approaches: Linear Regression, Bagged decision tree with Laplace smoothing, SVM and Proximal SVM (PSVM) over four industries manufacturing, wholesale, retail and services. Ryser \& Denzler \citep{ryser2009selecting} used k-fold cross validation procedure to compare the performance of different credit rating models. They concluded that although non-parametric approaches such as random forest, neural network and generalized additive models still outperform the parametric models, they may overfit during the training process. Hajek \& Michalak \citep{hajek2013feature} studied the performance of supervised machine learning techniques such as Multilayer Perceptron (MLP), Support Vector Machines (SVM), radial Basis Function Neural Network (RBF), Naive Bayes (NB), Random Forest (RF), Linear Discriminant Classifier (LDC) and Nearest Mean Classifier (NMC) on both US and European non-financial companies. They focused on 81 financial and non-financial variables of 852 US companies and 43 financial and non-financial variables of 244 European companies. They reduced the dimension of data by applying feature selection techniques such as wrapper algorithms. They concluded that US credit ratings are more affected by the size of companies and market value ratios while the European ratings depend more on profitability and leverage ratios.  Zhong et al. \citep{zhong2014comparing} provided a comparison study over reliability, overfitness and error distribution of different learning algorithms such as SVM, Back propagation neural network, Extreme Learning Machine (ELM) and Incremental Extreme Learning (IELM). They considered 6 financial variables on both S\&P and Moody's rating categories to classify the US corporate credit ratings.

Bahrammirzaee \citep{bahrammirzaee2010comparative} provided a comparative analysis of the application of three categories of techniques: artificial neural networks, expert systems and hybrid intelligent systems on credit evaluation. Their analysis categorized articles where artificial neural network outperform other methods as well as articles where neural network performs as same as or even worse than other traditional methods such as decision trees and linear regression. Dima et al. \citep{dima2016companies} compared the performance of neural networks with Bayesian approach on a large sample of companies ($3000$ companies) applying for credit to the same bank in Romania. Khemkhem et al. \citep{khemakhem2015credit} implemented neural networks and linear discriminant analysis to forecast Tunisian corporate credit ratings. They concluded neural networks outperform discriminant analysis. Hajek et al. \citep{hajek2014predicting} compared the performance of artificial immune classification algorithms (AIRS1, AIRS2, AIRS2-p and CSCA) and several well-known machine learning techniques (MLP, Decision tree, SVM and RBF) on investment and non-investment grade firms. Both groups of input variables were collected for 520 U.S. companies in the year 2010, 195 classified as investment grade and 325 as non-investment grade. They used accuracy and misclassification cost as performance measures of the studied classifiers. They also provided confusion matrix for each classifier (TP, TN, FP, FN). First, they conclude that AIRS2 outperforms AIRS1, AIRS2-p and CSCA on the test set based on the studied measures. They also show that MLP and Decision tree are the best models by providing the highest accuracy and the lowest misclassification cost, respectively.

Wallis et al. \citep{wallis2019credit} provided a comparative analysis of several most popular machine learning techniques (multinomial logistic regression, linear discriminant analysis, regularized discriminant analysis,artificial neural network, support vector machine, gaussian process classifier, random forest, gradient boosting machine) to predict Moody’s lone term credit rating. They studied these models on 308 of the S\&P 500 companies from January 2016 to November 2017. They concluded the top three performing models were random forest, artificial neural network and support vector machines where random forest outperforms the others with $64.6\%$ accuracy. 

Moscatelli et al. \citep{moscatelli2019corporate} applied random forest, gradient boosted trees, logistic regression, penalized logistic regression and linear discriminant analysis to predict default probability of Italian non-financial firms from 2011-2017. They studied economic and financial ratios as potential indicators of corporate defaults. They also added a set of credit behavioral indicators on the firm-bank relationship to the studied financial ratios. They concluded random forest outperform all other models in predicting default/non-default credit rating.

A summary of discussed approach including number of financial variables, input dataset, output data set provider and the corresponding accuracy is provided in tables below. Table \ref{tab:1} refers to ANN-based approach. Table \ref{tab:2} addresses SVM-based models and Table \ref{tab:3} represents Decision tree-based methods.  


\begin{table}[H]

    \centering
    
    \caption{Comparison results for ANN approach}
   \rotatebox{90}{
   \begin{small}
    \begin{tabular}{|c|c|c|c|c|c|}
    \hline
         Reference&Year&Num. of Input Var.&Input Data&Credit Rating Provider&Accuracy\\
         \hline
         \citep{garavaglia1991application}&1991&87&797 Companies&S\&P&85\%\\
         \hline
         \citep{moody1994architecture}&1994&10&196 Indudtrial Firms&S\&P&36\%\\
         \hline
         \citep{huang2004credit}&2004&5&36 Commercial Banks&S\&P&80.00\%\\
         \hline
         \citep{huang2004credit}&2004&6 \& 16&Taiwan Dataset&Taiwan Ratings Corporation&75.68\%\\
         \hline
         \citep{kim2005predicting}&2005&&1080 US companies&S\&P&84\%\\
         \hline
         \citep{brabazon2006credit}&2006&8&791 Non-financial US Firms&S\&P&85\%\\
         \hline
         \citep{cao2006bond}&2006&17&US Manufactoring Companies&S\&P&80.28\%\\
         \hline
         \citep{lee2007application}&2007&10&3017 Korean Companies&Korea Information Service, Inc&59.93\%\\
         \hline
         \citep{yu2008credit}&2008&13&Jap. Consumer Credit Card&Application Approvals&88.08\%\\
         \hline
         \citep{yu2008credit}&2008&12& 60 UK Corporations&UK Ratings&85.87\%\\
         \hline
         \citep{khashman2010neural}&2010&24&German Dataset& German Ratings&85.9\%\\
         \hline
         \citep{kim2010support}&2010&10&Korean SMEs&Korean Default Rates&64.23\%\\
         \hline
         \citep{hajek2012credit}&2012&52&852 US Mining Companies&S\&P&80\%\\  
         \hline
         \citep{hajek2012credit}&2012&14&154 US Municipalities&Moody's&71.62\%\\
         \hline
         \citep{west2000neural}&2000&24&German and Australian Data& German and Australian Rating&error:0.2243\\
         \hline
        \citep{khemakhem2015credit}& 2015 & 15 & 86 Tunisian companies & Tunisian credit rating & 82.55\%\\        
         \hline
         \citep{zhao2015investigation}&2015&20&German Credit Dataset& German Credit Ratings& 87\%\\
         \hline
         \citep{hajek2014predicting} & 2014 & 20 & Investment and non-investment grade & S\&P & 88.44\%\\
         \hline
         \citep{addo2018credit} & 2018 & 181 and 10 & 117019 companies asking for a loan& European Bank & RMSE: 0.044\\
         \hline
        \citep{daniel2019corporate}& 2019 & 17& Random US companies & S\&P and Moody's&83\%\\
         \hline
        \citep{wallis2019credit} & 2019 & 27 & 308 companies from S\&P 500 & Moody's & 63.6\%\\
         \hline
    \end{tabular}
    \end{small}
    } 
    \label{tab:1}
\end{table}

\begin{table}[H]
    \centering
    \caption{Comparison results for SVM approach}
    \rotatebox{90}{
    \begin{tabular}{|c|c|c|c|c|c|}
    \hline
         Reference&Year&Num. of Input Var.&Input Data&Credit Rating Provider&Accuracy\\
         \hline
         \citep{huang2004credit}&2004&14&36 Commercial Banks&S\&P&80.00\%\\
         \hline
         \citep{huang2004credit}&2004&6&Taiwan Dataset&Taiwan Ratings Corporation&79.73\%\\
         \hline
         \citep{cao2006bond}&2006&17&US Manufactoring Companies&S\&P&84.61\%\\
         \hline
         \citep{lee2007application}&2007&10&3017 Korean Companies&Korea Information Service, Inc&67.22\%\\
         \hline
         \citep{ye2008multiclass}&2008&33&Four US industries&Moody's&64\%\\
         \hline
         \citep{yu2008credit}&2008&13&Jap. Consumer Credit Card&Application Approvals&79.91\%\\
         \hline
         \citep{yu2008credit}&2008&12& 60 UK Corporations&UK Ratings&77.84\%\\
         \hline
         \citep{kim2010support}&2010&10&Korean SMEs&Korean Default Rates&66.16\%\\
         \hline
         \citep{hajek2011credit}&2011&5&852 US firms&S\&P&85.96\%\\
         \hline
         \citep{hajek2011credit}&2011&12&452 Czech municipals&local experts&89.76\%\\
         \hline
         \citep{kim2012corporate}&2012&14&Korea Dataset&Korea Ratings Corporation&67.98\%\\
         \hline
         \citep{yeh2012hybrid}&2012&18&2470 Taiwanese companies&Taiwanese Rating&74.4\%\\
         \hline
       \citep{hajek2014predicting} & 2014 & 20 & Investment and non-investment grade & S\&P & 87.28\%\\
         \hline
         \citep{pai2015credit}&2015&18&Taiwan Dataset&Taiwan Ratings Corporation&86\%\\
         \hline
         \citep{wallis2019credit} & 2019 & 27 & 308 companies from S\&P 500 & Moody's & 60.1\%\\
        \hline
    \end{tabular}
   }
    \label{tab:2}
\end{table}

\begin{table}[H]
    \centering
    \caption{Comparison results for Tree-based approach}
    \rotatebox{90}{
    \begin{tabular}{|c|c|c|c|c|c|}
    \hline
         Reference&Year&Num. of Input Var.&Input Data&Credit Rating Provider&Accuracy\\
         \hline
         \citep{shin2001case}&2001&12&Korean Companies&Korean Ratings&69.3\%\\
         \hline
         \citep{koh2004credit}&2004& & & & 74.2\%\\
         \hline
         \citep{ye2008multiclass}&2008&33&Manufacturing/Wholesale/Retail/Services&Moody's&47.1\%\\
         \hline
          \citep{hajek2014predicting} & 2014 & 20 & Investment and non-investment firms & S\&P & 86.13\%\\
         \hline
          \citep{pedro}&2016&33&Healthcare, Financial, Technology Sectors in USA&S$\&$P and Moody's&$>70\%$\\
         \hline
          \citep{addo2018credit} & 2018 & 181 and 10 & 117019 companies asking for a loan from a bank & European Bank & RMSE: 0.044\\
         \hline
        \citep{wallis2019credit} & 2019 & 27 & 308 companies from S\&P 500 & Moody's & 64.6\%\\
        \hline
        \citep{moscatelli2019corporate} & 2019 & 26 & Italian non-financial firms & Italian Credit Register & $> 84\%$\\
        \hline
    \end{tabular}
    }
    \label{tab:3}
\end{table}

\section{Analytical Methods}\label{sec:Methodology}
Supervised learning methods are based on learning from observations. Let $X \subset R^n$ and $Y$ be the input and output space respectively. The output space identifies the classification type . When $Y = \{-1,+1\}$, the problem is a binary classification. If $Y = \{1,2,..,m\}$, the problem is a multiple class classification. $Y \subset R^n$ leads to a regression type problems. During the learning process, the classifier assumes there exists a hidden function $Y = f(x)$ in a training set $L = \{(x_i,y_i)\}_{i = 1}^N$. The classifier attempts to find a prediction for this function such that the error is minimized.  

There are multiple ways to determine the exact nature of function $f$. In computer science and statistics these are generically named machine learning algorithms. We are focusing an a few specific machine learning algorithms described in this section. The previous section and the summary Tables \ref{tab:1}, \ref{tab:2}, and \ref{tab:3} detail these methods as the most popular in credit rating problem. This is the reason we chose them for implementation. 





\subsection{Bagged Decision Trees (BDT)}
Let $L$ be the learning/training set consists of $N$ instances. Bagging or bootstrap aggregation algorithm creates $B$ bootstrap samples where each sample consists of $n<N$ randomly chosen instances but with replacement from $L$. Each of these $B$ bootstrap samples train the same classifier which in this case is a decision tree. In order to predict an outcome for each instance of the testing set, the new case must be run down each of the $B$ decision trees. The prediction is obtained by a maximum number of votes among the $B$ classifiers.

This ensemble structure combines predictions of multiple classifiers to provide a better performance than unique learner. The combination of learners in bagging structure reduces the risk of classification error (variance) for those unstable high variance classifiers such as decision trees. 

In Bagged Decision Trees (BDT), the possibility of over-fitting of individual tree is less concerned. For this reason, the individual decision trees are grown deep and the trees are not pruned. Bagged Decision tree has only two hyper-parameters, the number of trees and the number of samples. However, they may have many similarities which leads to a high correlation in predictions. The structure of Random Forest will address this problem. A more detailed description of these methods can be found in \citep{breiman1984classification}.

\subsection{Random Forest (RF)}
Random Forest (RF) is another algorithm based on ensemble of classification trees which is developed by Breiman \citep{breiman1984classification,breiman2001random}. The only difference between RF and BDT is that RF takes one extra step. In addition to taking the random subset of data, it also chooses randomly a subset of $X$ at each node and calculate the best split at that node only within the given subset of $X$.  This structure provides uncorrelated or weakly correlated predictions.  Also, there is no pruning step, which means all the trees of the forest are grown deep. RF has only two hyper-parameters, the number of variables in the random subset at each node and the number of trees in the forest. 
Moreover, RF ranks variables by the importance of a variable based on the classification accuracy, while considering the interaction between variables.

\subsection{ANN: Multi Layer Perceptron (MLP)}
A perceptron is a linear classifier that separates two classes with a straight line. Let X be the input vector and y be a perceptron that produces a single output:
$$y = \Phi (\sum_{i=1}^n w_i x_i + b) = \Phi (w^Tx + b)$$
where w is a vector of input weights, b is a bias and $\Phi$ denotes a non-linear activation function.

A multilayer perceptron is a deep, artificial neural network composed of more than one perceptron. In this NN, the input layer receives the input vector, the output layer makes a decision and provide a prediction, and in between these two, there exists a number of hidden layers. The input signal moves forward from the input layer to the output layer, through the hidden layers. 
Let $l$ represents the number of input units, $m$ represents the number of hidden units and $n$ represents the number of output units. Each hidden layer unit is defined as:

$$h_j = f (\sum_{i\rightarrow j} w_{ij}^{(1)} x_i + w_{0j}^{(1)}) = f (w^{(1)}x)$$
where $f$ is an activation function on hidden layers, $i = \{1,2,\cdots,l\}$ and $j = \{1,2,\cdots,m\}$. The common choice for $f$ is a logistic function $f(z) = \frac{e^{z}}{1 + e^{z}}$. this hoice is important and depends on the nature of the data.

Similarly, the output units are denoted as:
$$y_k = g (\sum_{j\rightarrow k} w_{jk}^{(2)} h_j + w_{0k}^{(2)})$$
where $g$ is another activation function and $k = \{1,2,\cdots,n\}$. Again, the common choices are logistic functions. 

Considering the layers:
$$y_k(x,(w^{(1)},w^{(2)})) = g ( w_{0k}^{(2)} + \sum_{j=1}^m w_{jk}^{(2)} f (\sum_{i=1}^l w_{ij}^{(1)} x_i + w_{0j}^{(1)}))$$

The number of parameters required to build this neural network is :
$$l.m + m + m.n + n = (l+1)m + (m+1)n$$

So depending on the weights and the choice of activation functions, different values of outputs are obtainable.

When the decision of the output layer is computed against the real outputs, the error will be calculated. The parameters of the model, the weights and the bias terms, will be adjusted during the training in order to minimize the error.
$$\Delta w = \eta * e_k * x_{jk}$$
where $e_k$ represents the error on the output unit $k$ and $x_{jk}$ is the input that caused the error and $\eta$ is the learning rate which defines how much to change the weight to correct the error.

Partial derivative of the error w.r.t. the weights and biases are back propagated through the MLP. Therefore the change of weights in the hidden layer follows the previous formula except the error term which will be computed as: 
$$e_j = w_j * e_k * d(y_k)$$ 
where $d$ is the transfer derivative. The network keeps doing these forward and backward passes until the error can go no lower. This state is called convergence. 

\subsection{Support Vector Machines (SVM)}
SVM is an algorithm that implements non linear boundaries between classes by transforming the input data into a high dimensional space. This mapping into a new space is a task of kernel functions which make the input data set linearly separable. In the new space, SVM constructs a maximal margin hyperplane which provides a maximum separation between output classes. The training observations that are closest to the maximal margin hyperplane are called support vectors. It introduced first by Vapnik in a form of quadratic optimization problem with bound constrains and one linear equality constraints.
The standard SVM classifier is only applicable on binary type classification problems.

\[
  \begin{cases}
    w^T \Phi (x_i) + b \geq +1      & \quad \text{if } y_i = +1\\
    w^T \Phi (x_i) + b \leq -1  & \quad \text{if } y_i = -1\\
  \end{cases}
\]
such that $w$ and $b$ represent the weight vector and bias respectively. $\Phi (): R^n \rightarrow R^{n_k}$ represents a mapping function and $w^T \Phi (x_i) + b = 0 $ the optimal separating hyperplane. 

The formula above is equivalent to $$y_i [w^T \Phi (x_i) + b] \geq +1$$ for $\forall i = \{1,2,\cdots,N\}$

This formula constructs two hyperplanes on the opposite sides of the optimal hyperplane with the total margin size of $\frac{2}{||w||^2}$.

The classifier takes a decision then based on a decision function $$sgn(w^T \Phi (x_i) + b)$$
In order to solve the linearly non-separable problems, it is general to use a slack variable $\xi$ which allows a miss-classification error. The optimization problem containing the weighted miss-classification error will be formed as:

\begin{equation*}
 \begin{aligned}
 & \underset{w,b,\xi}{min}
 & & \frac{1}{2} w w^T + C\sum_{i=1}^N \xi_i\\
 & \text{subject to}
 & & \begin{cases}
    y_i [w^T \Phi (x_i) + b] \geq 1 - \xi_i      & \quad i = 1,2,\cdots,N\\
    \xi_i \geq 0  & \quad i = 1,2,\cdots,N\\
  \end{cases}
 \end{aligned}   
\end{equation*}
where $C \in R^+$ is a tuning hyperparameter that weights the importance of miss-classification errors. This optimization problem is solvable using Lagrangian where the Lagrangian multiplier $\alpha$ exists for each training observation. The non-zero $\alpha$'s represent the support vectors in the training set.

Therefore, the formula above leads to a dual problem:

\begin{equation*}
\begin{aligned}
& \underset{\alpha}{\text{max}}
& & \frac{1}{2} \alpha^T Q \alpha - e^T \alpha  \\
& \text{subject to}
& & \begin{cases}
    0 \leq \alpha_i \leq C      & \quad i = 1,2,\cdots,N\\
    y^T \alpha = 0  & \quad \\
  \end{cases}
\end{aligned}
\end{equation*}
where $e$ is the vector of ones, $Q_{ij} = y_i y_j K(x_i,x_j)$ is a $N*N$ positive semi-definite matrix and $K(x_i, x_j) = \Phi(x_i)^T \Phi(x_j)$ is a kernel function.
The choice of kernel functions depends on the type of the problem in hand. The linear kernel function is $K(x_i, x_j) = x_i^T x_j$, the radial basis kernel function is $K(x_i, x_j) = exp \{-\gamma ||x_i - x_j||^2\}$, the polynomial kernel function with degree $d$ is $K(x_i, x_j) = (\gamma x_i^T x_j +r)^d$ and the sigmoid kernel is $K(x_i, x_j) = tanh\{\gamma x_i^T x_j +r\}$

The final SVM classifier will be in a form of:
$$sgn \Big(\sum_{i}^N \alpha_i y_i K(x,x_i) + b\Big)$$

\subsection{Multi Class Support Vector Machines}
The conventional SVM explained above is a binary classifier. For multi-class classification problems, we need to modify its structure. There are different approaches that extends binary SVM structure to solve the multi-class classification problems, two of which have been explained and used in this study.

\subsubsection{One versus All Approach}
Consider an M-class classification problem. The problem can be decomposed into M binary subproblems. One-against-all approach constructs M binary SVM classifiers, each of which separates one class from the other M-1 classes.
Let N be the number of training samples $(x_1, y_1),(x_2,y_2), \dots, (x_N,y_N)$. The $i-$th SVM classifier is trained with all the training examples of the $i-$th class with positive labels, and all the others with negative labels \citep{liu2005one,kim2012corporate}.
\subsubsection{One versus One Approach}
This approach constructs binary SVM classifiers for all pairs of classes; in total, there are $\frac{M(M-1)}{2}$ pairs. For each given pair, binary SVM classifier follows the optimization problem above to maximize the margin between classes \citep{kim2012corporate}. 

\section{Research Methodology}

In order to have a fair assessment of the methodologies developed to predict credit rating, we need to have an equal evaluation footing for the methods. To do so we need to use the same dataset with the same structure of training, validation and testing. Second we need to have a fair measure for performance evaluation. We introduce a so called Notch measure in section \ref{sec:NotchDistance} that we believe it is appropriate for assessing the performance of credit rating algorithms. In our applications we focus on methods described in section \ref{sec:Methodology}. These are the methods used by the majority of credit rating literature (see Tables \ref{tab:1}, \ref{tab:2}, and \ref{tab:3}).

\subsection{Data sets}
\begin{figure}[H]
\caption{Distribution of credit ratings for each sector}
    \centering
    \includegraphics[width=1\linewidth, height=5cm]{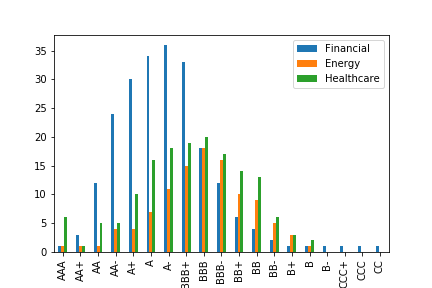}
    
    \label{fig:11}
\end{figure}

Previous Studies show that machine learning techniques are able to predict corporate credit ratings using their historical financial variables \citep{huang2004credit,hajek2012credit,pai2015credit}. In this study, we consider $52$ stocks of financial sector, $28$ stocks of energy sector and $44$ stocks of healthcare sector. The input data set covers these corporate historical financial variables from 1990-2018 for financial sector and from 2009 to 2018 for energy and healthcare sectors. These variables are taken from both Bloomberg and Compustat and have been merged together to reduce the number of missing values. Among all available variables, we selected $16$ financial variables for financial sector and $20$ variables for both energy and healthcare sectors in order to predict the credit ratings. The primary factors in the selection of these variables are the availability of data and the influence of a variable on the credit rating based on previous studies \citep{huang2004credit}. The output data which are the corresponded credit ratings are taken from S\&P and it contains $19$ rating classes starting from AAA to CC. These variables are listed in Tables \ref{tab:4},\ref{tab:5}. Table \ref{tab:6} and Figure \ref{fig:11} present the distribution of each sector with respect to rating classes.    
\begin{table}[H]
\caption{Financial variables used in financial sector}
    \centering
    \begin{tabular}{cc}
    \hline
         &Variables\\
         \hline
         \hline
         $V_1$&Asset/Equity\\
         
         $V_2$&Total Common Equity\\
         
         $V_3$&Total Asset\\
         
         $V_4$&Total Invested Capital\\
         
         $V_5$&Total Debt/Total Equity\\
         
         $V_6$&Total Debt/Total Asset\\
         
         $V_7$&Total Liabilities\\
         
         $V_8$&Sgort and Long term Debt\\
         
         $V_9$&Long Term Borrow\\
         
         $V_{10}$&Return on Asset\\
         
         $V_{11}$&Debt/Market Cap\\
         
         $V_{12}$&Operating Margin\\
         
         $V_{13}$&IS-OPER-INC\\
         
         $V_{14}$&Net Income\\
         
         $V_{15}$&Profit Margin\\
         
         $V_{16}$&EPS-FOR-RATIOS\\
         \hline
    \end{tabular}
    
    \label{tab:4}
\end{table}

\begin{table}[H]
\caption{Financial ratios used in energy and healthcare Sectors}
\centering  
    \begin{tabular}{ccc}
    \hline
         &Ratio \\
         \hline
         \hline
         $R_1$&Debt/EBITDA\\
         
         $R_2$&FFO/Total Debt\\
         
         $R_3$&EBITDA/Interest\\
         
         $R_4$&FFO/Interest\\
         
         $R_5$&CFO/Debt\\
         
         $R_6$&FFO/Net Profit\\
         
         $R_7$&NWC/Revenue\\
         
         $R_8$&Current Asset/Current Liabilities\\
         
         $R_9$&(FFO+Cash)/Current Liabilities\\
         
         $R_{10}$&EBITDA/Revenues\\
         
         $R_{11}$&Cash/Total Debt\\
         
         $R_{12}$&Total Debt/Tangible Net worth\\
         
         $R_{13}$&Total Debt/Revenue\\
         
         $R_{14}$&Debt/Capital\\
         
         $R_{15}$&Cash/Asset\\
         
         $R_{16}$&Total Fixed Capital/Total Fixed Assets\\
         
         $R_{17}$&Equity/Asset\\
         
         $R_{18}$&NWC/Total Assets\\
         
         $R_{19}$&Retained Earnings/Total Assets\\
         
         $R_{20}$&EBITDA/Total Assets\\
         \hline
    \end{tabular}
    
    \label{tab:5}
\end{table}

         
         
         
         
         
         
         
         
         
         
         
         
         
         
         
         
        
         
         
         
    

\begin{table}[H]
\small
\caption{Number of Companies in each class of rating on December 2018}
    \centering
    \begin{tabular}{cccc}
    \hline
         Rating Classes& Financial Sector & Energy Sector & Healthcare Sector \\
         \hline
         \hline
         AAA&1&1&6\\
         
         AA+&3&1&1\\
         
         AA&12&1&5\\
         
         AA-&24&4&5\\
         
         A+&30&4&10\\
         
         A&34&7&16\\
         
         A-&36&11&18\\
         
         BBB+&33&15&19\\
         
         BBB&18&18&20\\
         
         BBB-&12&16&17\\
         
         BB+&6&10&14\\
         
         BB&4&9&13\\
         
         BB-&2&5&6\\
         
         B+&1&3&3\\
         
         B&1&1&2\\
         
         B-&1&&\\
         
         CCC+&1&&\\
        
         CCC&1&&\\
         
         CCC-&&&\\
         
         CC&1&&\\
         \hline
         
    \end{tabular}
    
    \label{tab:6}
\end{table}

\subsection{Evaluation Methodology}\label{sec:NotchDistance}

The typical way to evaluate a machine learning model is the standard accuracy as the proportion of observations correctly classified in the test data. For example, the column ``Accuracy'' in Tables \ref{tab:1}, \ref{tab:2}, and \ref{tab:3} present a summary of these percentages in prior work. A quick inspection of these numbers appear to say that accurate predictions based on publicly available financial variables using machine learning techniques is achievable. However, predicting credit rating is a problem that goes beyond most other classification problems. 

First, the credit rating data is temporal. That is, a prediction has to be done at a point in time and data from future quarters may not be used in prediction. Most techniques divide data into 80\% training and 20\% test data sampled at random which means invariably that future observations will be used in the model development. We will investigate the models performance when the test data is made of future observations. 

Second issue is that corporate credit ratings do not change very often. Due to this, in about $90\%$ of observations the credit rating does not change with respect to the previous quarter and in only $10\%$ of observations the credit rating changes. Therefore, creating a simple model that just predicts the same credit rating as the previous quarter will have a 90\% accuracy! This beats the majority of the results in Tables \ref{tab:1}, \ref{tab:2}, and \ref{tab:3}. 
Thus, we believe it is crucial to investigate the performance of the classifier on those observations where the rating is changing from the previous quarter. 

Third issue is that the proportion of observations correctly classified tells us nothing about how far from the true rating were the inaccurate predictions. We quantify the distance between predictions and real ratings using what is known in literature as \textit{Notches} distance. To create a numerical value of the distance we sort the ratings from best (AAA) to worst (CC). We give a numerical value to each rating (e.g., AAA is $1$)

\begin{definition}[Notch Distance]\label{definition:Notch}
To formulate this distance, we let $y$ denote the true rating of an observation. Let $\hat y$ denote the prediction given by a particular model. Then define a random variable: Notch Distance as $Y= \hat y - y$. To asses the distribution of this variable for a particular model that makes predictions $\hat y_1,\ldots,\hat y_N$,  we calculate $F$ the \textit{frequency} of the notch $i$ value:
$$\text{F}(i) = \sum_{k \in N} I_{(\hat{y}_k - y_k = i)} /N$$   
where $N$ is the total number of observations in the test set, $i \in \{\cdots,-1,0,1,\cdots\}$, and $I$ is the indicator function taking value $1$ every time the condition is satisfied and $0$ otherwise.
\end{definition}

In the definition above $F(0)$ is the proportion of observations predicted correctly. 
To create a measure of accuracy we can use the random variable created. Specifically, we can calculate the expected value: 
$$DC = \mathbf E[Y]= \sum_{i}i * \text{F(i)}$$
and we may regard it as a \textit{Dissimilarity Coefficient}. The numerical value is the expected notch distance corresponding to a particular method. In a similar way we may define an \textit{Absolute Dissimilarity Coefficient}
$$ADC =\mathbf E[|Y|] = \sum_{i} |i| * \text{F(i)}$$
We can of course calculate the standard deviation of the variable $Y$ and that would be a measure of how far the typical prediction is from the actual rating.

\begin{theorem}[Jensen's Inequality]
Let $X$ be a random variable with finite expectation, and let $g: R \rightarrow R$ be a convex function whose domain includes the range of X. Then,
$$E[g(X)] \geq g(E[X])$$
If the function $f$ is strictly convex, then the inequality holds with equality if and only if $X$ is degenerate.

As a consequence of Jensen's inequality, let $X=|Z|$ and $g:x \rightarrow x^2$. Then,
$E[g(X)]=E(Z^2)$ and $g(E[X])=E[|Z|]^2$
hence Jensen's inequality reads $$E[|Zt|]^2 \leq E[Z^2] = \sigma^2$$.
\end{theorem}

Both these measures will be influenced by the percentage of the correct predictions (value $0$). To have a more accurate measure in terms of average notch distance when the prediction is wrong we can calculate a conditional expectation:
$$\mathbf E[Y\mid Y\neq 0] = \sum_{i\neq 0} i * \frac{\text{F(i)}}{\sum_{j\neq 0 }\text{F(j)}}$$
This conditional expectation eliminates the correct prediction and its expectation should be a better measure of how many notches we expect the algorithm to be off when the prediction fails.

\section{Experiment Results and Analysis}

\begin{figure}[H]
    \centering
    \includegraphics[scale = 0.45]{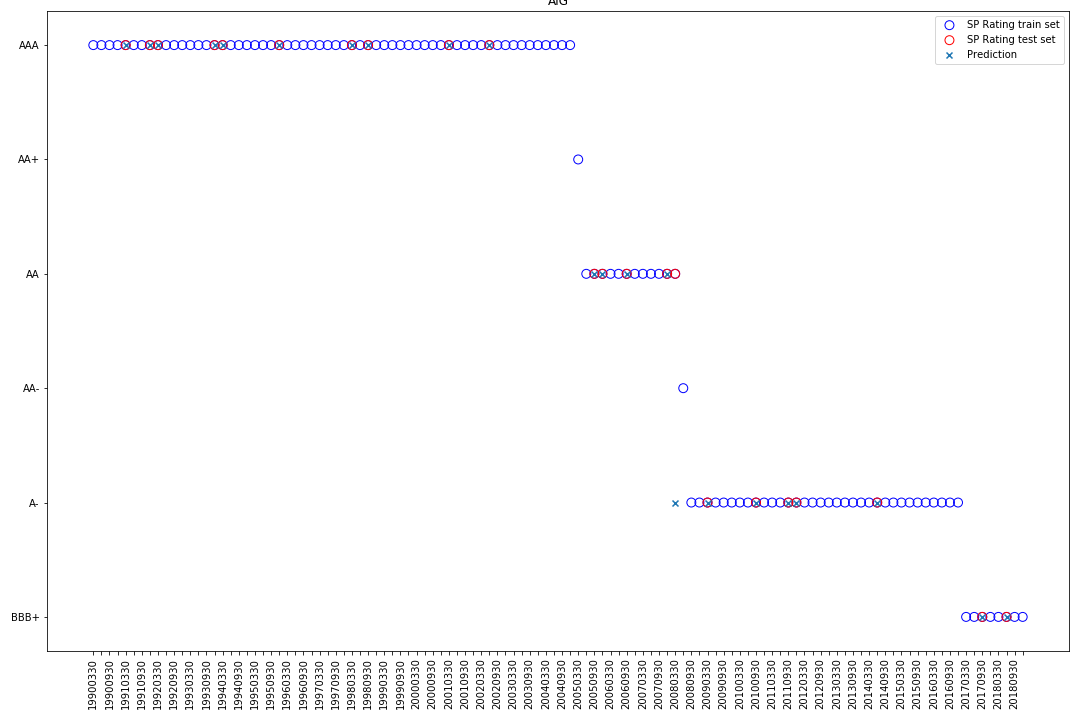}
    \caption{Train set, Test set and Predictions of AIG using Bagged Decision Tree}
    \label{fig:1}
\end{figure}

\begin{figure}[H]
    \centering
    \includegraphics[scale = 0.45]{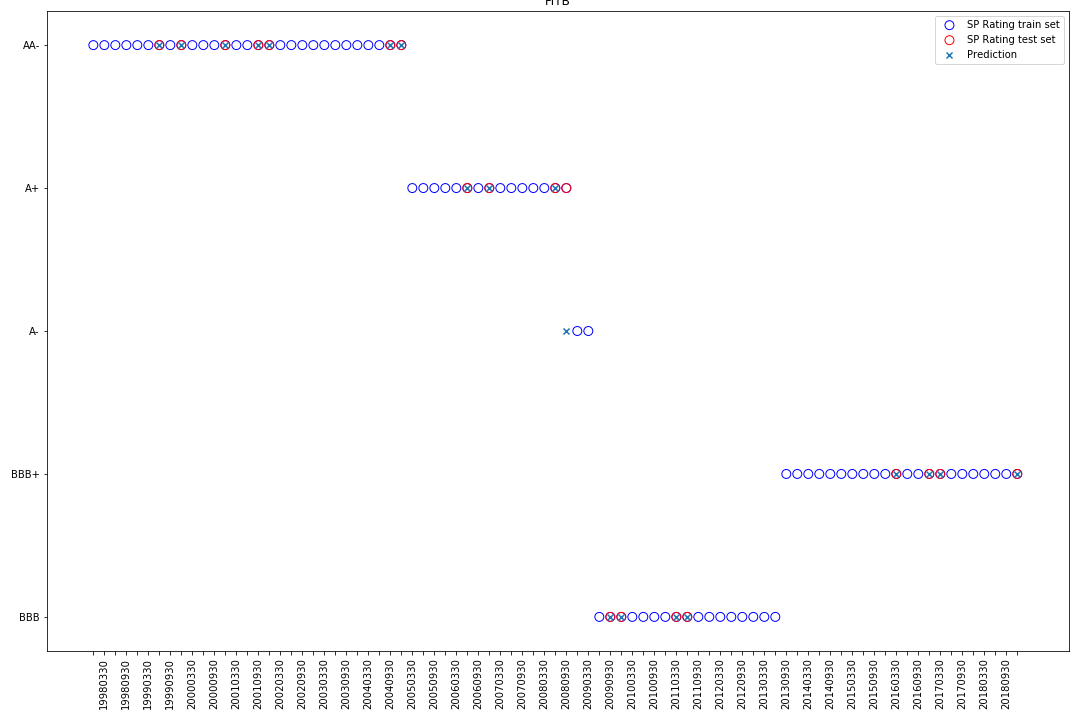}
    \caption{Train set, Test set and Predictions of FITB using Random Forest}
    \label{fig:2}
\end{figure}

\begin{figure}[H]
    \centering
    \includegraphics[scale = 0.45]{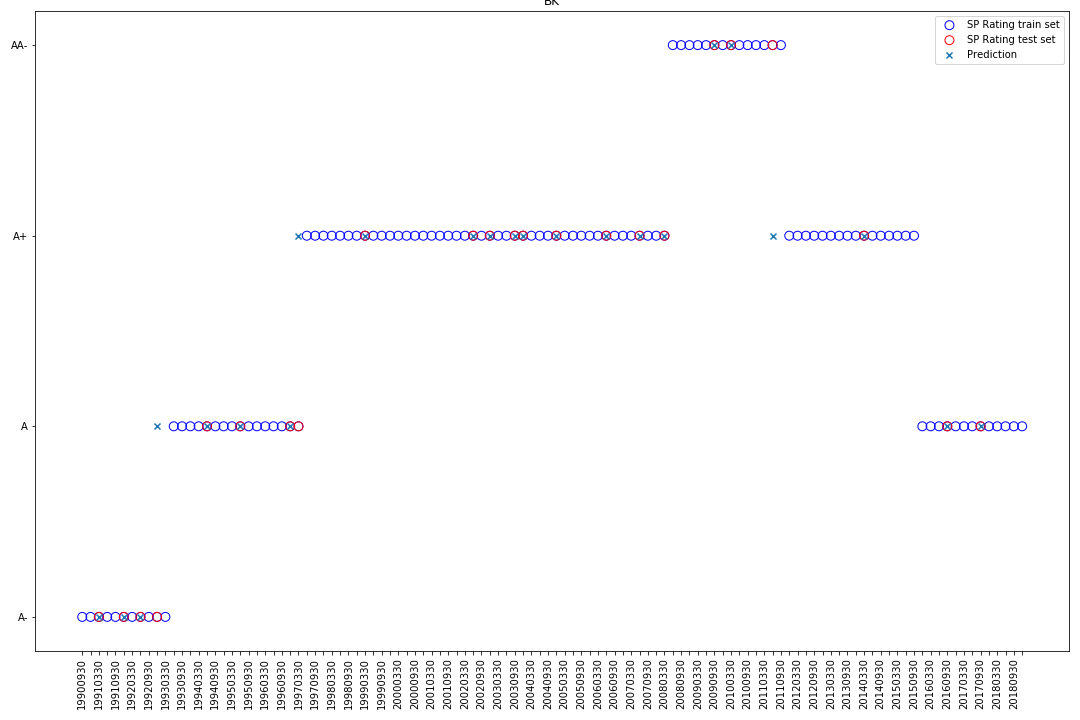}
    \caption{Train set, Test set and Predictions of BK using MLP}
    \label{fig:3}
\end{figure}

In this section we study the performance of four Machine learning techniques: Bagged Decision Tree (BDT), Random Forest (RF), Multiple Layer Perceptron (MLP) and Support Vector Machine (SVM) techniques to predict corporate credit ratings. Figures \ref{fig:1},\ref{fig:2},\ref{fig:3} illustrate some examples of the type of results obtained.  Figures depict the training set (blue circle), test set (red circle) and predictions (green x) for various stocks for three methods: Bagged Decision Tree, Random Forest, and MLP.

\subsection{Analysis of Prediction Accuracy for different Machine Learning models}

To address overfitting and obtain a more accurate results, we use a, so called, 10-fold cross validation procedure. Specifically, each sector data is split into three subsets, a training set, a validation set and a test set. For example, for financial sector dataset, a training set of $70\%$, which is $4220$ data points, a validation set of $10\%$ which is $603$ data points and a test set of $20\%$ which is $1206$ data points of input data have been considered. Table \ref{table:1} presents prediction accuracy of each of these techniques based on 10-fold cross validation procedure. The SVM technique is a binary type classification technique. However, in our case we have several categories that need to be classified not just two. Thus we use two different SVM-based methods to classify the rating categories \citep{kim2012corporate}.

\begin{table}[H]
\begin{tabular}{|c|ccc|}
\hline
     & Financial Sector& Energy Sector & Healthcare Sector\\
     \hline
     \hline
     Bagged Decision Tree & 84.21\%&82.11\%&83.90\%\\
     
     Random Forest & 82.83\%&84.45\%&82.97\%\\
     
     Multi Layer Perceptron & 73.95\%&78.19\%&76.63\%\\
     
     One Vs. One-SVM & 42.12\%&75.31\%&71.29\%\\
     
     One Vs. All-SVM &40.14\% & 59.17\%& 61.89\%\\
     \hline
\end{tabular}
\caption{Accuracy of different techniques}
\label{table:1}
\end{table}

Table \ref{table:1} contains the type of results that are reported by the majority of articles applying Machine Learning algorithms to the credit rating problem. According to the results we obtain it is clear that Bagged Decision Tree and Random Forest outperformed MLP and SVM-based approaches. This type of performance was observed for all datasets and periods considered. Based on the Tables \ref{tab:1} and \ref{tab:2}, most of the related studies on S\&P credit scores, prove that neural network and support vector machine obtain more than 80\% prediction accuracy on US firms. Only one study obtained 36\% accuracy of neural network on industrial firms \cite{moody1994architecture}. 

For a financial analysis it is important to have a measure of how far the prediction is when the algorithm fails. To this end recall we introduced the Notch Distance in Definition \ref{definition:Notch} on page \pageref{definition:Notch}. 

A summary of the distribution of the notch distance $Y$ is presented in Table \ref{table:3}. As expected values for $F(0)$ are exactly the prediction accuracy percentages in Table \ref{table:1}.

\begin{table}[H]
\caption{Computed Notches for all stocks in three sectors}
\begin{tabular}{|c|c|ccc|}
\hline
     Sector & Methods & Zero &One Absolute &Greater than One Absolute \\
     \hline
     \hline
     Financial & Bagged Decision Tree & 84.21\% & 12.01\% & 3.78\% \\
     
     & Random Forest & 82.83\% & 12.08\% & 5.09\% \\
     
     & Multi Layer Perceptron & 73.95\% & 15.92\% & 10.13\% \\
     
     & One Vs. One-SVM & 42.12\% & 32.13\% & 25.75\% \\
     
     & One Vs. All-SVM& 40.14\%&32.28\%&27.58\%\\
     \hline
     Energy & Bagged Decision Tree & 82.11\% & 13.98\% & 3.91\%\\
     
     &Random Forest & 84.45\% & 11.85\% & 3.70\%\\
     
     &Multi Layer Perceptron & 78.19\% & 15.48\% & 6.33\% \\
     
     &One Vs. One-SVM & 75.31\% & 16.98\% & 7.71\%\\
     
     &One Vs. All-SVM & 59.17\% & 20.64\% & 20.19\%\\
     \hline
     Healthcare & Bagged Decision Tree & 83.90\% & 13.51\% & 2.59\% \\
     
     &Random Forest & 82.97\% & 11.74\% & 5.29\% \\
     
     &Multi Layer Perceptron & 76.63\% & 16.08\% & 7.29\% \\
     
     &One Vs. One-SVM & 71.29\% & 16.03\% & 12.68\%\\
     
     &One Vs. All-SVM & 61.89\% & 24.50\% & 13.61\%\\
     \hline
     
\end{tabular}

\label{table:3}
\end{table}

\begin{table}[H]
\caption{Expected Values and Standard Deviation of all notches in three sectors}
\centering
\begin{tabular}{|c|c|ccc|}
\hline
     Sector & Methods &$E[X]$&$\sigma(X)$&$E[|X|]$ \\
     \hline
     \hline
     Financial& Bagged Decision Tree & 0.0501 & 0.7614 & 0.2719 \\
     
     & Random Forest & 0.0068 & 0.7028 & 0.2731 \\
     
     & Multi Layer Perceptron & 0.1381 & 1.0152 & 0.5114  \\
     
     & One Vs. One-SVM & 0.2661 & 1.6917 & 1.4584  \\
     
     & One Vs. All-SVM & 0.6263 & 1.4562 &1.4699 \\
     \hline
     Energy & Bagged Decision Tree & 0.0980 & 0.5817 & 0.2851 \\
         
     & Random Forest & -0.0594 & 0.5892 & 0.3212 \\
         
     & Multi Layer Perceptron & -0.0800 & 0.8421 & 0.3858 \\
         
     & One Vs. One-SVM & -0.0685 & 0.7224 & 0.3481 \\
         
     & One Vs. All-SVM & -0.0066 & 0.9117 & 0.6491  \\
     \hline
     Healthcare & Bagged Decision Tree & 0.1975 & 0.3512 & 0.2959 \\
         
     & Random Forest & -0.2154 & 0.6317 & 0.3223 \\
         
     & Multi Layer Perceptron & -0.0531 & 0.6854 & 0.4486 \\
         
     & One Vs. One-SVM & 0.1191 & 0.7515 & 0.6531 \\
         
     & One Vs. All-SVM & 0.2461 & 0.9683 & 0.8372  \\
     \hline
\end{tabular}

\label{table:4}
\end{table}

Table \ref{table:4} provides expected value and standard deviation of the notch distance for every model. The columns represent the three sectors we are analyzing: financial sector, energy sector and healthcare sector. First we look at the expected value as an indication of symmetry of prediction. The standard deviation and expected absolute value are measure of how far the prediction is from the actual rating. 

The results in the table generally indicate that Bagged decision tree and Random Forest methods have a better performance compared to the ANN and SVM. The largest proportion of predictions is accurate thus the notches distance is zero. We next calculate the conditional expectation of the notches measurement value when we are wrong. Table \ref{table:7} shows this conditional expected values and the conditional standard deviation.

\begin{table}[H]
\caption{Conditional Expected Values and Standard Deviation of all notches in three sectors}
    \centering
    \begin{tabular}{|c|c|ccc|}
    \hline
         Sector & Methods &$E[X|X\neq0]$&$\sigma(X|X\neq0)$&$E[|X||X\neq0]$\\
         \hline
         \hline
         Financial & Bagged Decision Tree &-0.0018 & 0.8619&1.2795\\
         
         & Random Forest&-0.0489 & 0.9037 & 1.1451\\
         
         & Multi Layer Perceptron &0.2391&1.3294&1.3912\\
         
         & One Vs. One-SVM &0.0911&1.8011&2.0594\\
         
         & One Vs. All-SVM & 0.7216&1.4991 & 1.9171  \\
         \hline
         Energy & Bagged Decision Tree &0.0770 & 0.3735 & 0.8264 \\
         
         &Random Forest& -0.0885 & 0.3010 & 0.7587 \\
         
         &Multi Layer Perceptron & -0.1317 & 0.5615 & 0.9991 \\
         
         &One Vs. One-SVM & -0.2415 & 0.3512 & 0.9387 \\
         
         &One Vs. All-SVM &-0.5131 &0.4411 & 1.2853 \\
         \hline
         Healthcare & Bagged Decision Tree & 0.1973 & 0.2218 & 0.5313 \\
         
         &Random Forest & -0.4516 & 0.3054 & 0.6214 \\
         
         &Multi Layer Perceptron & -0.2712 & 0.3614 & 0.7778 \\
         
         &One Vs. One-SVM & -0.0710 & 0.5784 & 1.0496 \\
         
         &One Vs. All-SVM & 0.4052 & 0.6413 & 1.2384 \\
         \hline
    \end{tabular}
    
    \label{table:7}
\end{table}
 
Looking at the standard deviation and expected absolute for the unconditional Notches distance in Table \ref{table:4} it would appear that the performance for all sectors is roughly similar. However, looking at the conditional numbers in Table \ref{table:7} we see that healthcare and perhaps energy sectors are predicted better than the financial sector. The numerical values in Table \ref{table:7} also provide information about how far the prediction is when it is wrong. For example the Bagged decision tree is on the average $0.81$ notches away from the true value.  
 
\subsection{Comparing the performance of the various Machine Learning Credit Ratings with the differences between different credit rating agencies.}

In the previous section we compared the accuracy of the prediction of Machine learning models with the corporate ratings provided by Standard and Poor's, Moody's and Fitch.  These three credit agencies do not always agree on the ratings of companies. Therefore, the variation of their ratings as measured by Notch distance is useful in order to assess a measure of an acceptable standard deviation of a specific credit rating. 

In Table \ref{tab:ratingAgencies} we present the same statistics as in the previous table using quarterly ratings from each of the rating agencies. We can see that the notch distance between the rating agencies shows less agreement between themselves than most of the Machine learning methods considered in this paper. This should give some confidence in the machine learning methods in this paper.

\begin{table}[]
\caption{Notch distances for the three rating agencies S$\&$P, Moody's, and Fitch quarter\label{tab:ratingAgencies}}
    \centering
    \begin{tabular}{|c|cccccc|}
    \hline
         & $E[X]$ & $\sigma(X)$ & $E[|X|]$ & $E[X|X\neq0]$ & $\sigma(X|X\neq0)$ & $E[|X||X\neq0]$ \\
         \hline
         \hline
         S$\&$P and Moody's & 0.33 & 1.23 & 0.76 & 0.54 & 1.53 &  1.43 \\
         S$\&$P and Fitch & -0.05 & 0.95 & 0.64 & -0.11 & 1.32 &  1.23 \\
         Moody's and Fitch & -0.30 & 1.02 & 0.75 & -0.51 & 1.28 & 1.27 \\
         \hline
    \end{tabular}
    
    \label{tab:my_label}
\end{table}

\subsection{Capturing rating changes}

Generally, credit ratings do not change very much from quarter to quarter. Thus for example if a dataset contains 90\% of quarters in which ratings did not change from the previous quarter, a vary naive model that would predict the same rating as the one from the previous quarter would have 90\% accuracy! Thus, the most important issue when predicting credit rating based on information available is about predicting a rating change when the actual rating will change. Table \ref{table:10} shows the percentage of credit rating changes that have been captured by each model. These numbers are actually much lower than the precision of the models. They are larger than we expected, however, the performance of the Bagged decision tree and random forest is not that far from the MLP and SVM in capturing rating changes. These ideas will be investigated in future work as predicting changes is the most important part of the credit analysis. 

\begin{table}[H]
\caption{Captured Changes}
\centering
\begin{tabular}{|c|ccc|}
\hline
     & Financial Sector & Energy Sector & Healthcare Sector\\
     \hline
     \hline
     Bagged Decision Tree & 67.64\% & 88.88\% &77.27\%\\
     
     Random Forest & 71.56\% & 77.77\% & 83.33\%\\
     
     Multi Layer Perceptron & 53.92\% & 77.77\% &66.66\%\\
     
     One Vs. One-SVM &  29.41\% & 77.77\%&58.33\%\\
     
     One Vs. All-SVM &25.49\% & 33.33\% &41.66\%\\
     \hline
\end{tabular}

\label{table:10}
\end{table}

\section{Conclusion and Future Directions}
In this work we first provide a survey of current literature using machine learning techniques to predict corporate credit ratings. The survey pointed toward support vector machine (SVM) and artificial neural network (ANN) as methods that produce the most successful results. We implemented these methods and two other methods based on decision trees which in addition have the ability to select best features for the classification problems. We applied all methods for the same datasets spanning 2009 to 2018 for companies from  two different sectors: energy, and healthcare and another dataset spanning 1990 to 2018 for companies from financial sector. All our results indicate that decision tree methods are in fact outperforming SVM and MLP. To check the results we introduced a Notches Distance concept. This measure helps us quantify how far the predictions are from the true values when the algorithm fails to produce the correct prediction.
Using this new measure the results were unchanged. Since this is a new measure we need to have a benchmark for the results we obtained. To this end we calculated the Notch Distance using ratings provided by S\&P, Moody's and Fitch on the same companies. We found that the best algorithms produce a Notch Distance from the true ratings which is comparable with the distance between ratings produced by different rating agencies on the same company.

\bibliographystyle{apalike}
\bibliography{references}
\end{document}